\documentclass[prx,twocolumn,aps,nofootinbib,floatfix,superscriptaddress]{revtex4-1}
\usepackage{amsmath}
\usepackage{amssymb}
\usepackage{graphicx}
\usepackage{verbatim}
\usepackage[colorlinks=true,linkcolor=blue,citecolor=blue,urlcolor=blue]{hyperref}
\usepackage{txfonts}
\usepackage{xr}

\begin{document}

\title{Continuously parametrized quantum simulation of molecular electron transfer reactions}
\author{Frank Schlawin}
\thanks{present address: Max-Planck-Institute for the Structure and Dynamics of Matter, Hamburg, \href{mailto:frank.schlawin@mpsd.mpg.de}{frank.schlawin@mpsd.mpg.de}}
\affiliation{Clarendon Laboratory, University of Oxford, Parks Road, Oxford OX1 3PU, United Kingdom}
\affiliation{Physikalisches Institut, Albert-Ludwigs-Universit\"at Freiburg, Hermann-Herder-Stra\ss e 3, 79104 Freiburg, Germany}

\author{Manuel Gessner}
\email{manuel.gessner@ens.fr}
\affiliation{Laboratoire Kastler Brossel, ENS-Universit\'e PSL, CNRS, Sorbonne Universit\'e, Coll\`ege de France, 24 Rue Lhomond, 75005 Paris, France}
\affiliation{Physikalisches Institut, Albert-Ludwigs-Universit\"at Freiburg, Hermann-Herder-Stra\ss e 3, 79104 Freiburg, Germany}

\author{Andreas Buchleitner}
\email{abu@uni-freiburg.de}
\affiliation{Physikalisches Institut, Albert-Ludwigs-Universit\"at Freiburg, Hermann-Herder-Stra\ss e 3, 79104 Freiburg, Germany}
\affiliation{Freiburg Institute for Advanced Studies, Albert-Ludwigs-Universit\"at Freiburg, Albertstra\ss e 19, 79104 Freiburg, Germany}
\affiliation{Erwin Schr\"odinger International Institute for Mathematics and Physics, University of Vienna, Boltzmanngasse 9, 1090 Vienna, Austria}

\author{Tobias Sch\"{a}tz}
\affiliation{Physikalisches Institut, Albert-Ludwigs-Universit\"at Freiburg, Hermann-Herder-Stra\ss e 3, 79104 Freiburg, Germany}
\affiliation{Freiburg Institute for Advanced Studies, Albert-Ludwigs-Universit\"at Freiburg, Albertstra\ss e 19, 79104 Freiburg, Germany}

\author{Spiros S. Skourtis}
\affiliation{Department of Physics, University of Cyprus, P O Box 20537, 1678 Nicosia, Cyprus}
\affiliation{Freiburg Institute for Advanced Studies, Albert-Ludwigs-Universit\"at Freiburg, Albertstra\ss e 19, 79104 Freiburg, Germany}

\date{\today}

\begin{abstract}

A comprehensive description of molecular electron transfer reactions is essential for our understanding of fundamental phenomena in bio-energetics and molecular electronics.
Experimental studies of molecular systems in condensed-phase environments, however, face difficulties to independently control the parameters that govern the transfer mechanism with high precision.
We show that trapped-ion experiments instead allow to reproduce and continuously connect vastly different regimes of molecular charge transfer through precise tuning of, e.g., phonon temperature, electron-phonon interactions, and electronic couplings.
Such a setting allows not only to reproduce widely-used transport models, such as Marcus theory. It also provides access to transfer regimes that are unattainable for molecular experiments, while controlling and measuring the relevant observables on the level of individual quanta.
Our numerical simulations predict an unconventional quantum transfer regime, featuring a transition from quantum adiabatic- to resonance-assisted transfer as a function of the donor-acceptor energy gap, that can be reached by increasing the electronic coupling at low temperatures.
Trapped ion-based quantum simulations thus promise to enhance our microscopic understanding of molecular electron transfer processes, and may help to reveal efficient design principles for synthetic devices. 

\end{abstract}

\maketitle

\section{Introduction}
Molecular electron transfer (ET) reactions are the fundamental steps in many chemical and biological processes \cite{marcusreview,kuzul,advchem,maykuhn,balz,Nitzan}. A detailed understanding of the underlying physics is crucial in such diverse fields as catalysis \cite{balz}, photosynthesis \cite{blank}, biological signalling \cite{cambridge}, molecular electronics \cite{cuevas} and energy conversion~\cite{currop,Scholes2017}. 
Marcus theory,  the fundamental theory for thermally activated nonadiabatic ET rates, has been developed already in 1956 \cite{marcusreview}, and has since been generalized to include quantum or classical vibrations that modulate the  energy gap or the coupling \cite{marcusreview,kuzul,advchem,maykuhn,balz,Nitzan,annrev}. However, many ET reactions cannot be described by incoherent nonadiabatic rate theories. Currently, intensive research efforts are dedicated to non-standard transport regimes~\cite{currop,blum,acrecent}. These include fast charge and energy transfer reactions on the timescales of vibrational relaxation and electronic dephasing, such as the primary reactions in photosynthesis~\cite{currop}, nonergodic and nonequilibrium vibrational dynamics~\cite{blum,molphys}, the transition from incoherent to coherent transport~\cite{Scholes19}, the participation of quantum vibrations in promoting ultrafast ET, and the control of ET by external fields \cite{molphys}.

Understanding the particulars of a molecular ET reaction means deciphering how the molecular assembly and dynamics determine the dominant transfer mechanism and the magnitude of the transfer rate. 
A common experimental approach is to chemically modify the molecular and/or solvent structure with the aim of changing some of the fundamental parameters that determine the transport mechanism (e.g., tunneling barrier heights, electron-phonon coupling strengths, etc) \cite{advchem,winklgray,blum,acrecent}. To achieve tunable and independent control of these parameters is challenging, as best exemplified by the experimental efforts to verify Marcus theory:
The theory predicts that the logarithmic ET rate has an inverse-parabolic dependence on the electron donor-to-acceptor (D-A)  energy gap ($\Delta E = E_A - E_D$),  if all 
other parameters are kept constant. 
As the energy gap is increased, the rate increases (normal region), reaches a maximum (activationless transfer) and then decreases (inverted region). The first experimental verification of the inverted region was accomplished almost thirty years after its theoretical prediction~\cite{Miller84, Closs88} and led to the Nobel Prize for ET theory in 1992~\cite{NobelLecture}. In this seminal experiment, 
eight discrete, distinct values of the energy gap were realised by synthesizing eight different acceptor species, attached to a donor by a rigid molecular bridge. 
Obviously, despite its spectacular success, such a strategy is unable to monitor the competition between distinct ET mechanisms, under continuous tuning of the system parameters which control the subtle interplay between direct electronic couplings and strong interactions with phononic baths. 

An experimental platform which precisely offers this kind of continuous tunability are arrays of electrostatically interacting ions trapped by harmonic external potentials \cite{WinelandBlatt2008,Haeffner2008,Blatt12,Schaetz12}
with an accuracy of individual parameter control evidenced by their performance among the best atomic clocks \cite{NISTPRL2017}. These systems are currently considered among the leading architectures for the realization of quantum simulators and computers \cite{Haeffner2008,Blatt12,Schaetz12,Tomi14,Bermudez2017}. 
The relevant coupling strengths, energy scales, and dissipative processes are determined with high accuracy by laser- and radiofrequency fields. The versatility of trapped-ion arrays renders them ideal candidates to study elementary quantum transport processes under controlled conditions. Previous studies have focused on the transport of electronic (effective spins) and vibrational excitations (phonons) across ion chains~\cite{BermudezPRL,Ramm, Abdelrahman,Bermudez16}, and on the influence of spin-phonon couplings on coherent exciton transport processes~\cite{Haeffner18}. 

Here we show that analog trapped-ion quantum simulations with strong electron-phonon interactions 
allow to continuously map out the parameter space of a paradigmatic ET scenario. We introduce and discuss a minimal setting readily accessible for state-of-the-art experiments that is able to reproduce the predictions of Marcus theory, including the nonadiabatic inverted regime. We further establish how lowering the temperature allows to reach the nonadiabatic quantum transport regime, where vibronic (induced by the strong coupling of {\em vibr}ational and electr{\em onic}
degrees of freedom) resonances determine the transfer rates.
Finally, we demonstrate that the ion-trap platform makes unconventional ET regimes accessible that are not observed in molecular systems.
We predict the crossover from conventional, nonadiabatic molecular ET to a quantum transfer regime where rates are limited by phonon lifetimes in the normal regime and heavily modulated by resonances in the inverted regime. These modulations can be understood as a consequence of trapped excited-state populations that do not participate in the adiabatic transport and only contribute to the transport on resonance. 

Our paper is structured as follows.
We start with an overview of the Hamiltonian and the phenomenology of ET in section~\ref{sec:ETtheory}. This is followed in section~\ref{sec:trappedionsetup} by an outline of the elementary experimental platform for proof-of-principle quantum simulations of ET models. 
In sections~\ref{sec:classicaltransport} and \ref{sec:quantumtransport}, we then present simulations of this platform, where we demonstrate that it can access well-established molecular ET regimes as well as exotic new transport regimes which are not observed in molecular samples. 
We then conclude with an outlook on possible extensions in section~\ref{sec:extensions} and a summary in section~\ref{sec:conclusions}.

\begin{figure}[tb]
\centering
\includegraphics[width=0.35\textwidth]{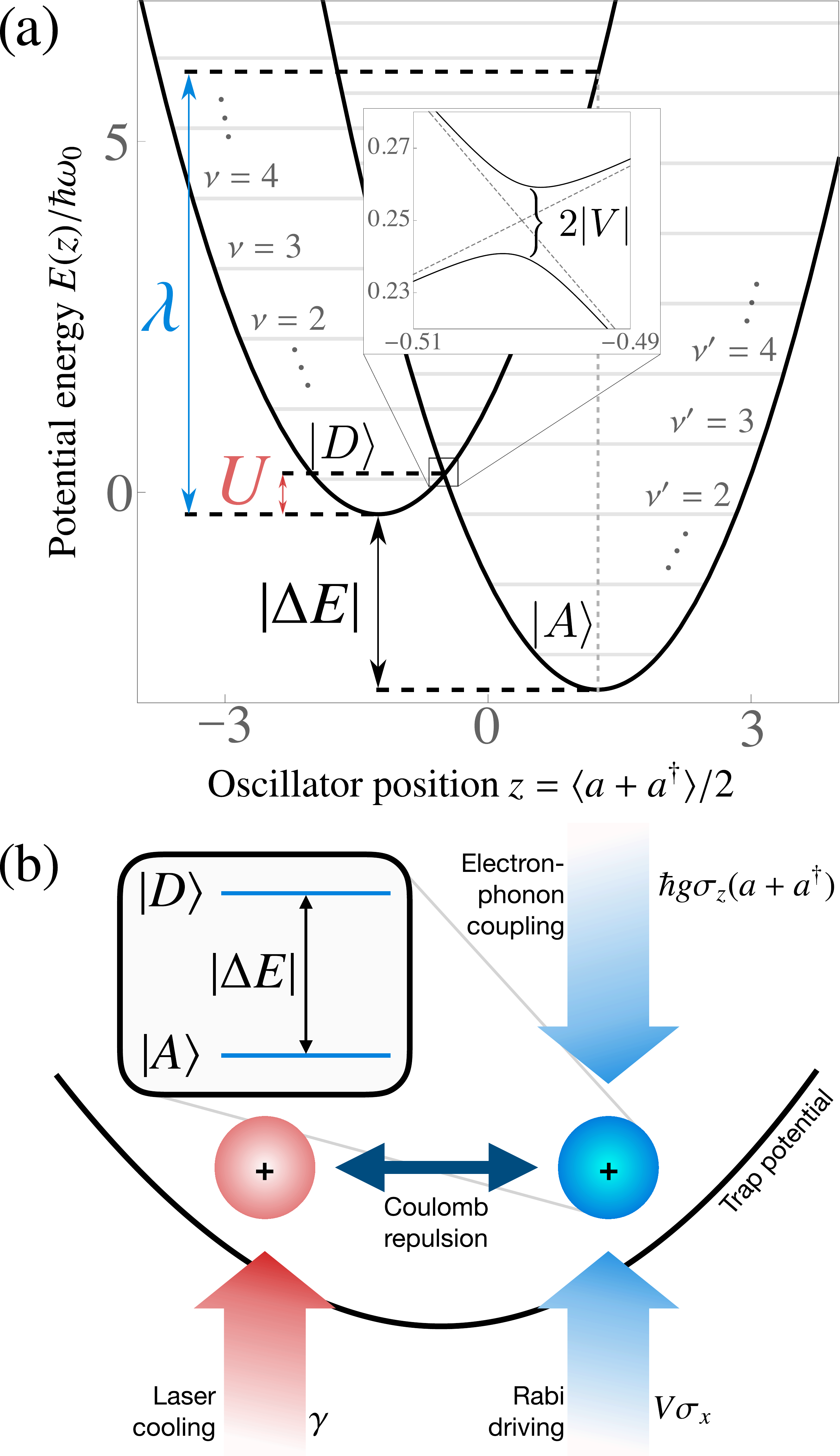}
\caption{
Electron transfer theory.
(a) Vibrational energy landscape deduced from the Hamiltonian~(\ref{eq.H_tot}), composed of the adiabatic Born-Oppenheimer (BO) surfaces for the electronic states $D$ and $A$, as a function of the reaction coordinate defined by the dimensionless harmonic oscillator position variable $z$. 
The diabatic BO surfaces, $E_{D}(z)$ (left parabola) and $E_{A}(z)$ (right parabola), respectively, are identified by dashed grey lines in the inset. For the weak coherent coupling assumed here, adiabatic and diabatic surfaces are indiscernible on the scale of the main plot.
The {\it D-A} energy gap $\Delta E=E_A-E_D$ [which is negative given the electronic level structure depicted in (b) below], the reorganization energy $\lambda$, the activation energy $U$, and the coherent coupling $V$ (inset) are indicated. Horizontal lines indicate vibrational levels $\nu$ ($\nu'$) of the {\it D} ({\it A}) BO surface upon quantization. Employed parameters are $|\Delta E| = 2.5 \hbar \omega_0$, $V = 0.01 \hbar \omega_0$, and $g = 2.5 \omega_0$, where $\omega_0$ denotes the (effective) trap frequency. 
(b) A `system' ion (blue) simulates the donor-acceptor system with two the electronic states $\vert D \rangle$ and $\vert A \rangle$ (akin to a spin $1/2$).
Laser-driven Rabi oscillations (blue upward arrow) induce coherent coupling of strength $V$ between donor and acceptor state. 
A second laser (blue downward arrow) induces spin-phonon couplings between the `system' ion's electronic degree of freedom and a collective motional mode of 
both ions. This mode is sympathetically cooled by laser interaction (red arrow) through a `cooling' ion (red) of a different species.
}
\label{fig.setupnew}
\end{figure}

\section{Electron transfer theory}
\label{sec:ETtheory}

ET from a donor (D) to an acceptor state (A) necessarily incorporates a large change of molecular charge distributions, such that the process is strongly coupled to the phononic degrees of freedom of the ET molecule. In its simplest form, such a process can be described by the Hamiltonian
\begin{align}
H &= \hbar\omega_0 a^{\dagger} a - \frac{\Delta E}{2} \sigma_z + V \sigma_x + \frac{\hbar g}{2} \sigma_z \left( a + a^{\dagger} \right)  \ \, \label{eq.H_tot}
\end{align}
where the first term describes a phononic mode which facilitates the ET. It gives rise to a quadratic potential in the reaction coordinate $z = (a + a^\dagger)/2$.
The second term, with $\sigma_z = |D\rangle\langle D| - |A\rangle\langle A|$, adds a state-dependent energy shift $\pm\Delta E/2$ of the vibrational harmonic oscillator potentials associated with the acceptor and donor levels, leading to a total energy offset $\Delta E$ (the donor-acceptor energy gap of the ET reaction) between donor and acceptor Born-Oppenheimer (BO) surfaces. Here, we adopt the definition $\Delta E = E_A - E_D$, which follows the convention of ET theory, where a negative $\Delta E$ expresses the fact that energy is released during the ET process \cite{marcusreview,kuzul,advchem,maykuhn,balz,Nitzan}. The third term coherently couples the donor and acceptor states, through $\sigma_x = |A\rangle\langle D| + |D\rangle\langle A|$. It has the strongest impact on the vibrational energy landscape at the degeneracy of the BO surfaces of the donor and the acceptor [see Fig.~\ref{fig.setupnew}~(a)], where it induces an avoided crossing of size $2|V|$, associated with a coherent population transfer at Rabi frequency $V/\hbar$. 
Finally, the fourth (coupling) term proportional to $\sigma_z(a+a^\dagger)=2\sigma_z z$, contributes a linear potential in the reaction coordinate $z$, with a positive or negative sign depending on whether the system resides in the donor or acceptor state $|D\rangle$ or $|A\rangle$. This displaces the harmonic oscillator vibrational potential minimum from the origin to $z_1=-g/2\omega_0$ (donor) and $z_2=g/2\omega_0$ (acceptor), respectively.

Crucially, the vibrational mode is subject to fast dissipation caused by the coupling to environmental modes. The dynamics is therefore described by the master equation 
\begin{equation}
\partial \rho(t)/\partial t = -(i/\hbar)[H,\rho(t)]+\mathcal{D}(\rho(t)), \label{eq.open}
\end{equation}
where the Lindblad dissipator is given by a damped harmonic oscillator \cite{Breuer}
\begin{align}
\mathcal{D} (\rho)&= \gamma (1+\bar{n})\left( a \rho a^{\dagger} - \frac{1}{2} \{ a^{\dagger} a, \rho \} \right) + \gamma\bar{n} \left( a^{\dagger} \rho a - \frac{1}{2} \{ a a^{\dagger}, \rho \} \right), \label{eq.cooling-rate}
\end{align}
and $\{A,B\}=AB+BA$.

To understand the ET process from donor to acceptor ($\vert D \rangle \rightarrow \vert A \rangle$) governed by the open-system dynamics ~(\ref{eq.open}), it is most intuitive to consider the potential landscape as a function of the reaction coordinate $z$, depicted in Fig.~\ref{fig.setupnew}~(a), with the BO surfaces $E_{D}(z)$ and $E_{A}(z)$. 
In molecular ET systems the reaction coordinate is the displacement of a molecular vibrational mode (or a set of modes) whose equilibrium conformation is changed by the electronic transition from $|D\rangle$ to $|A\rangle$. 
A key parameter of the transport process is the reorganization energy $\lambda$. It is the energy required to distort the minimum-energy conformation of the molecule in the electronic state $|D\rangle$ to the minimum-energy conformation in the state $|A\rangle$, without switching state from $|D\rangle$ to $|A\rangle$. It is thus the energy needed to displace the reaction coordinate on the donor BO surface from its minimum position to that of the acceptor BO surface~\cite{marcusreview,kuzul,advchem,maykuhn,balz,Nitzan}. In the  geometry determined by Eq.~(\ref{eq.H_tot}), this amounts to $\lambda = \hbar \omega_0 (z_1 - z_2)^2 =  \hbar g^2 / \omega_0$. 
A large reorganization energy compared to the phonon frequency signifies that such displacement will be associated with strong vibrational excitations. The phonon mode is thus strongly coupled to the  
electronic degree of freedom, and actively contributes to the ET reaction (i.e., it is ET-active). 
This is in contrast to models of vibrationally-assisted transport of excitations, where the coupling to phonons is typically less pronounced, and the transport is predominantly coherent and therefore reversible~\cite{cambridge}. 

Irreversible population transfer from $|D\rangle$ to $|A\rangle$ is mediated by classical or quantum mechanical mechanisms, depending on the dominant energy scale. 
These are the thermal phonon energy $k_BT$, the phonon energy $\hbar\omega_0$, the coupling strength $V$, the energy gap $\Delta E$, and the reorganization energy $\lambda$. On the one hand, the ratio of temperature and phonon energy determines whether the system is operated in a quantum or classical parameter regime. 
On the other hand, the relative (as compared to reorganization and phonon energy, respectively) strength of the avoided crossing of size $2|V|$ between the two BO surfaces controls the transition from nonadiabatic (between two distinct, diabatic BO surfaces) to adiabatic (between the minima of the same, lower BO surface) transfer. In this paper, we always consider cases where the initial and final electronic states are well localized in different regions of the molecule, known as localized polaron states with $|V| < \lambda/4$ \cite{marcusreview,kuzul,advchem,maykuhn,balz,Nitzan}. 

\section{Trapped-ion setup}
\label{sec:trappedionsetup}

\begin{table*}
\begin{tabular}{ l || l | l }
  \textbf{Molecular electron transport} & \textbf{Ion trap simulation} & \textbf{Parameter range}\\
  \hline			
  Reaction energy gap $\Delta E$ & Zeeman splitting $\Delta E$ & $|\Delta E|/\hbar \sim B\times 5-25$\,kHz$/\mu$T \cite{Langer}\\
  & & where $B\lesssim 30$\,mT \\
  Reorganization energy $\lambda$ & State-dependent dipole force $g$ $\Rightarrow$ $\lambda = \hbar g^2 / \omega_0$ & $g \lesssim 2\pi\times 500$\,kHz \cite{Govinda}\\
  Phonon energy $\omega$ & Trap frequency $\omega_0$ & $\omega_0\sim 2\pi\times 0.05-10$\,MHz \cite{Leibfried03,Haeffner2008,Schaetz12,Urabe} 
  \\
  &or laser detuning $\delta$ &$100\,\mathrm{Hz}\lesssim\delta \lesssim \omega_0/2$~\cite{Haeffner18}\\
  Electronic coupling $V$ & Rabi frequency $V / \hbar$
  & $V / \hbar \lesssim 2\pi\times 1$\,MHz \cite{Haeffner2008,Schaetz12}\\
  Resonance width $\Gamma$ & Sympathetic cooling rate $\gamma$ & $\gamma\lesssim 20$\,kHz \cite{Leibfried03,EIT2016} 
  \\
  Thermal energy $k_BT$ & Average phonon number $\bar{n}$ $\Rightarrow$ $k_B T = \hbar\omega_0/\ln[(\bar{n}+1)/\bar{n}]$ & $\bar{n}\gtrsim 0.01$ \cite{Leibfried03,Poyatos,Myatt,Cormick13,Barrett}
\end{tabular}
  \caption{Correspondence between characteristic parameters of electron transfer (left) and their counterparts in trapped-ion simulations (center), together with typical experimental values in ion-trap platforms (right). The notation in the left column is adopted from \cite{Nitzan}.
  }
  \label{tab.correspondence}
\end{table*}

The ion trap setup which we propose to simulate ET is sketched in Fig.~\ref{fig.setupnew} (b). Two positively charged ions are electrostatically trapped in a harmonic potential and repel each other by the Coulomb force. By assuming only small displacements around the equilibrium positions, the motion of the two-ion compound can be described by  $6$ normal phonon modes (2 in each dimension)~\cite{Leibfried03}. The electronic quantum state of the ion can be controlled with resonant electromagnetic fields. Employing near-resonant fields whose detuning is chosen to match a resonance condition involving the phonon modes allows us to control the coupling between select electronic and vibrational degrees of freedom.

For the implementation of ET physics, the first, `system', ion (blue dot) emulates the donor and acceptor states $\vert D \rangle$ and $\vert A \rangle$, respectively,  
by means of two long-lived electronic levels \cite{Haeffner2008,Blatt12,Schaetz12,Leibfried03}. These levels form an effective pseudospin, described by $-(\Delta E/2)\sigma_z$, where $\Delta E$ is the associated resonance frequency that can be controlled by an external magnetic field via the Zeeman effect. Driving the transition between these two levels with a resonant electromagnetic field induces Rabi oscillations, described by $V\sigma_x$, where $V/\hbar$ is the Rabi frequency [blue upward arrow in Fig.~\ref{fig.setupnew}~(b)]. The second, `cooling' ion (red dot) is used exclusively to sympathetically cool the collective motion of the two-ion compound \cite{Leibfried03, SympCooling,Blinov,Barrett,Drewsen,Home,Willitsch}. Choosing spectrally distinguishable ions or employing tightly focused laser beams prevents the cooling laser light from affecting the internal states of the system ion. The Hamiltonian of the relevant normal phonon mode of frequency $\omega_0$, determined by the trap potential~\cite{Leibfried03,James}, is given by $\hbar\omega_0a^{\dagger}a$, and the laser cooling evolution gives rise to the dissipator~(\ref{eq.cooling-rate}). Finally, the spin-phonon interaction term $\hbar g \sigma_z (a + a^{\dagger}) / 2$ between vibrational and electronic degree of freedom [blue downward arrow in Fig.~\ref{fig.setupnew}~(b)] can be realized by bichromatic laser driving with suitably chosen detuning close to the trap frequency~\cite{Kim,Porras08,Govinda,Haeffner18,LemmerNJP18}. The electron-phonon coupling constant $g$ is then determined via the laser intensities and detunings~\cite{Leibfried03, Haeffner18, LemmerNJP18}. An implementation with traveling-wave light is also possible. It requires a rotating-frame description of the dynamics, which causes the trap frequency $\omega_0$ to be replaced by the laser detuning. Throughout this article, we express all energy and time scales in units of $\omega_0$, which may represent an effective trap frequency. A similar spin-boson interaction term, where the coupling is given by $\sigma_x$ rather than $\sigma_z$, appears in the Rabi model that can also be simulated with high flexibility in a trapped ion setup using related methods~\cite{Lv18}. Taken together, the effective dynamics of the electronic degree of freedom of the system ion, together with the collective vibrational degree of freedom of the compound, is formally identical to the standard models of ET theory in Eq.~(\ref{eq.H_tot}).

All relevant aspects of the dynamics can be discussed and experimentally demonstrated starting with two trapped ions, coupled by the Coulomb force while sharing the same trap potential. 
Extensions to larger numbers of ions or different trap geometries and dimensionalities~\cite{Hakelberg} provide access to more complex situations which involve a larger number of electronic or phononic modes, giving rise, e.g., to intermediate bridge states~\cite{advchem2}. 

In contrast to real molecular systems, all these relevant system parameters are continuously and precisely tunable over wide intervals. We summarize the correspondence between the most important ET parameters and their analogs in the proposed trapped-ion simulation in Tab.~\ref{tab.correspondence}. In the following, we explore different transfer regimes by controlling individually the thermal phonon energy $k_BT$, the (effective) trap frequency $\omega_0$, and the coupling strength $V$ in numerical simulations of the time evolution governed by Eq.~(\ref{eq.open}).

\section{Classical nonadiabatic transport}
\label{sec:classicaltransport}
The classical regimes involve ET-active vibrational frequencies with $\hbar \omega_0 \ll \lambda$ and temperatures with $\hbar \omega_0 \ll k_BT$. Many molecular ET 
reactions in condensed phases (i.e., in solution, cellular or molecular-junction environments rather than in vacuum) belong to this category because low frequency environmental  
vibrations couple to ET. In this regime the Landau-Zener parameter 
$
\gamma_{\rm LZ} = \pi^{3/2} |V|^2 / ( \hbar \omega_0 \sqrt{\lambda k_B T} ) 
$
distinguishes the nonadiabatic (weak electronic coupling, $\gamma_{\rm LZ} \ll 1$) and adiabatic (strong electronic coupling, $\gamma_{\rm LZ} \gtrsim 1$) limits \cite{Sumi}. 

\begin{figure}[tb]
\centering
\includegraphics[width=0.4\textwidth]{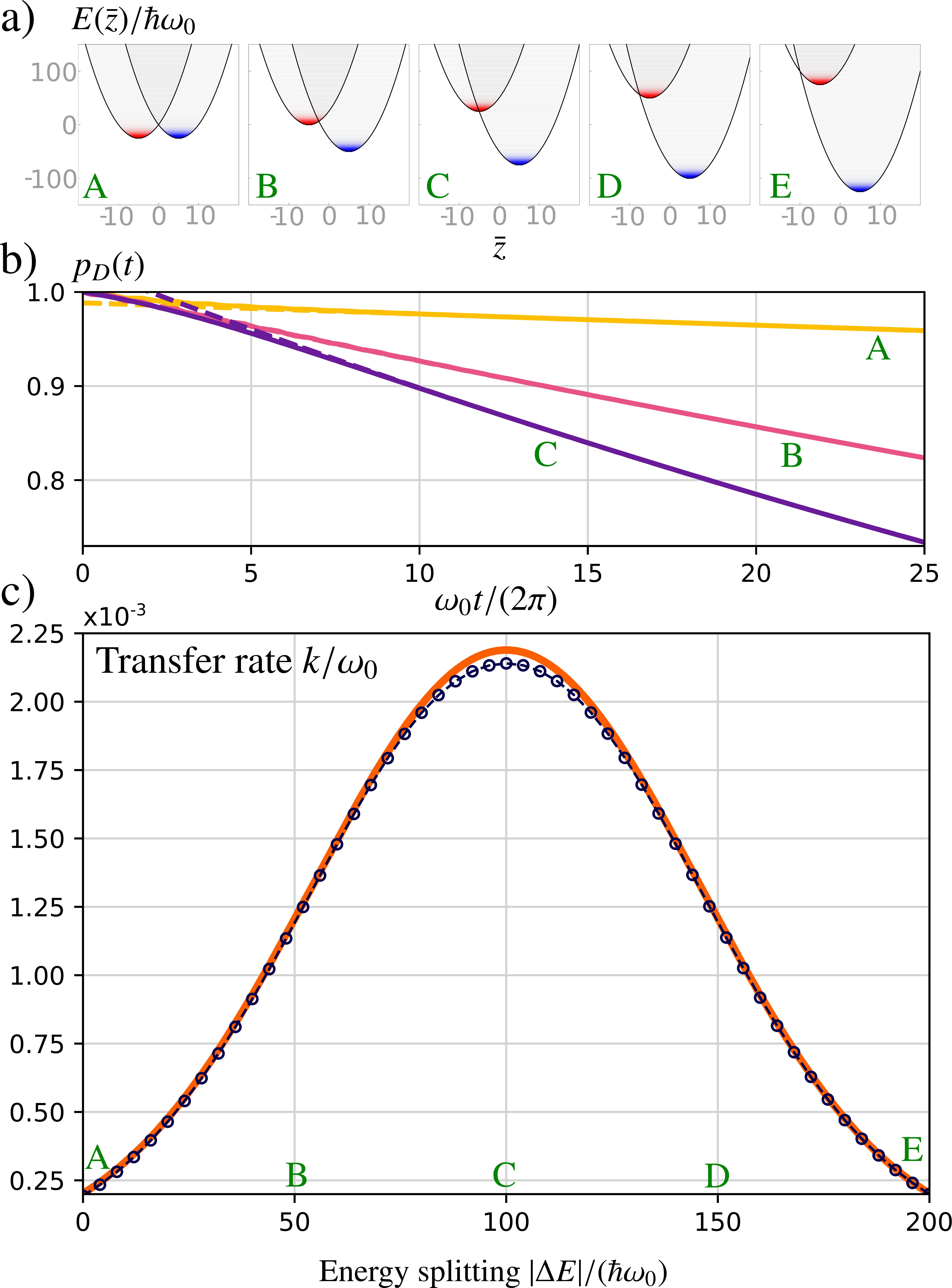}
\caption{{Classical nonadiabatic transfer.} (a) Diabatic Born-Oppenheimer energy surfaces and their vibrational levels in the classical nonadiabatic regime. The parameters are chosen as $V = 0.2 \hbar\omega_0$, $g = 10 \omega_0$,
$\gamma = 0.4 \omega_0/2\pi$, 
$\overline{n} = 10$, corresponding to $k_BT\approx 10.5\hbar\omega_0$ and $\lambda=100\hbar\omega_0$. 
The panels show $|\Delta E| = 0 \hbar \omega_0$ (A), $|\Delta E| = 50\hbar \omega_0$ (B), $|\Delta E| = 100\hbar \omega_0$ (C), $|\Delta E| = 150\hbar \omega_0$ (D), and $|\Delta E| = 200\hbar \omega_0$ (E). Vibrational states are indicated by horizontal grey lines, and red (donor) and blue (acceptor) shaded areas indicate the thermal phonon populations.
(b) 
Exponential fit (dashed lines) of the time evolution of the 
donor state population $p_D(t)$ (continuous lines) as generated by Eqs.~(\ref{eq.H_tot}) and~(\ref{eq.cooling-rate}), for the parameters of panels A (yellow), B (pink), and C (violet) in (a). (c) Transfer rates extracted from the exponential fits to $p_D(t)$ as a function of $|\Delta E|$ (dark blue circles). 
The orange line is the prediction of classical nonadiabatic transport theory, according to Marcus' formula~(\ref{eq.classical-Marcus}).}
\label{fig.cl-nonAd-regime}
\end{figure}

The classical nonadiabatic case (CN) is described by Marcus theory. Classical, thermally activated vibrations can tune the electronic states $|D\rangle$ and $|A\rangle$ to resonance, such that ET can take place at the crossing point of the two diabatic BO surfaces, while preserving the total 
energy of the system. The ET rate is then given by the expression~\cite{marcusreview,kuzul,advchem,maykuhn,balz,Nitzan}
\begin{align}
k_{\rm CN} &= \frac{2\pi}{\hbar} \frac{\vert V \vert^2}{\sqrt{4 \pi \lambda k_B T}} \exp \left(-\frac{U}{k_B T} \right), 
\label{eq.classical-Marcus}
\end{align}
where $U=(\Delta E + \lambda)^2 / (4 \lambda)$ is the activation energy to the resonance conformation on the donor BO surface [see Fig.~\ref{fig.setupnew}~(a)].
This transfer rate is proportional to $|V|^2$ and reaches a maximum in the activationless limit ($U =0$), when $| \Delta E | = \lambda$, such that the acceptor BO surface intersects the minimum of the donor surface [see Fig.~\ref{fig.cl-nonAd-regime}~(a), panel C].
The dependence of $\ln k_{\rm CN}$ on  the detuning $\Delta E$ is an inverted parabola---the seminal Marcus parabola, cf. the orange line in Fig.~\ref{fig.cl-nonAd-regime}~(c).

This parameter regime can be reached in the trapped-ion system for a thermal phonon population with a relatively low $\omega_0$. 
For our numerical simulations of the trapped-ion dynamics, we initialize the system at time $t = 0$ in $\vert D \rangle \langle D \vert \otimes \sum_{\nu} p_{\nu} \vert \nu \rangle \langle \nu \vert$, 
where $|\nu\rangle$ is a Fock state of $\nu$ phonons, weighted by a thermal distribution $p_{\nu}$ at temperature $T$. We extract the ET rate $k$ by fitting an exponential decay $a\exp(-kt)$ 
to the time evolution of the donor state population $p_D(t)$ after a transient evolution of $\omega_0 t/2\pi=5$ that is dominated by coherent dynamics (see also Appendix~\ref{app.decay-rate}). Typical evolutions are displayed in Fig.~\ref{fig.cl-nonAd-regime}~(b) for $k_B T\simeq 10.5\hbar  \omega_0$ 
and  $\gamma_{\rm LZ} \simeq 10^{-3}$. The corresponding average phonon number of $\bar{n}=10$ can be reached without requiring efficient cooling.
Figure~\ref{fig.cl-nonAd-regime}~(c) shows the extracted transfer rates from the trapped-ion simulation (dark blue circles). These follow nearly perfectly the Marcus prediction (orange line), Eq.~(\ref{eq.classical-Marcus}), which we plot without adjustable parameters. Both the normal and the inverted transport regimes can be simulated. Small deviations near the activationless case may be due to a broadening of the phonon resonances caused by the laser cooling process that is not captured by Marcus theory~(\ref{eq.classical-Marcus}).

\section{Quantum regimes}
\label{sec:quantumtransport}
\begin{figure}[tb]
\centering
\includegraphics[width=0.45\textwidth]{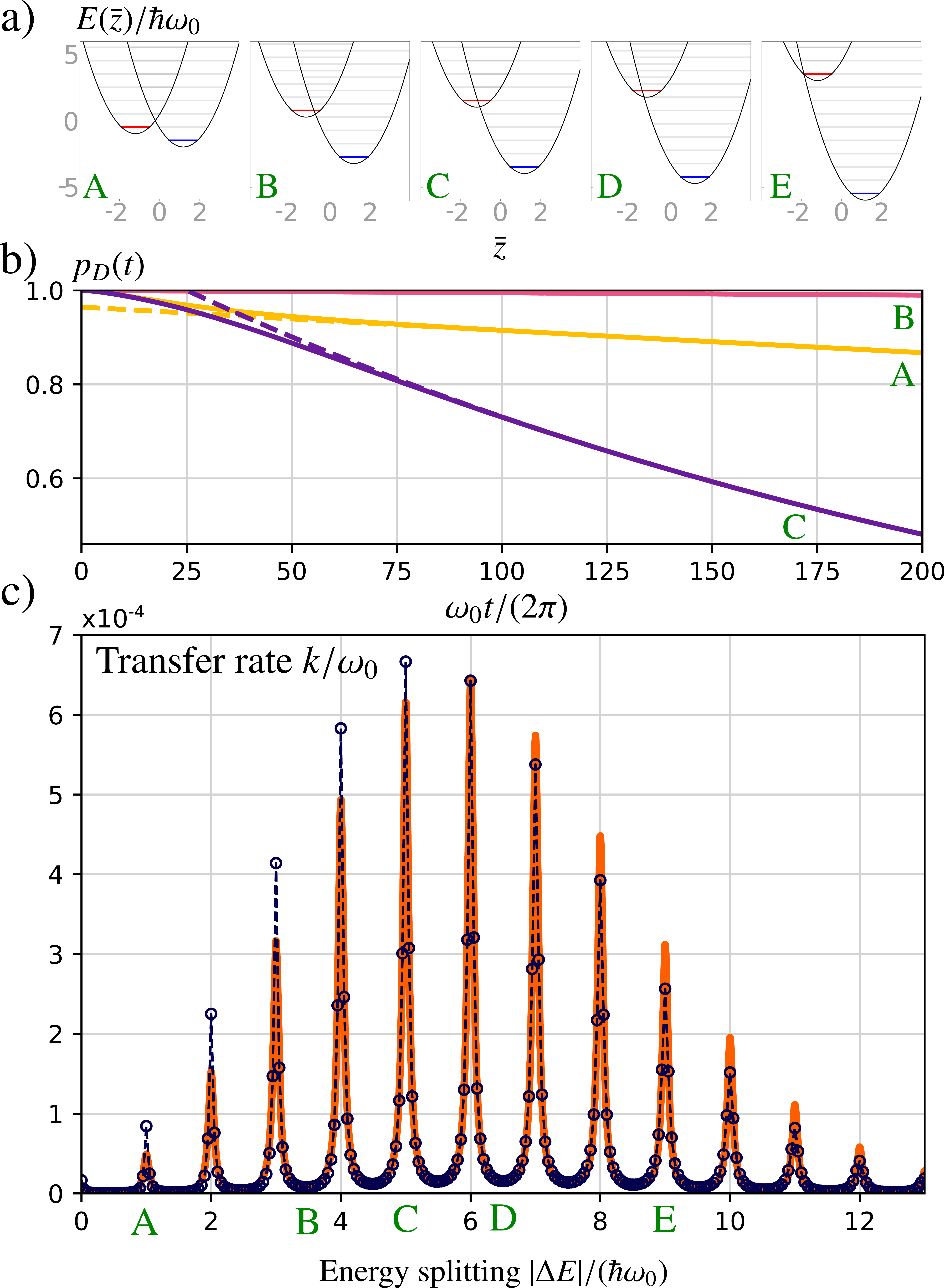}
\caption{(Color online) 
{
Quantum nonadiabatic (QN) transfer.} (a) Diabatic Born-Oppenheimer energy surfaces and their vibrational levels for the quantum nonadiabatic transfer regime. The system parameters are 
set to $V = 0.01 \hbar \omega_0$, $g = 2.5 \omega_0$, $\gamma=0.05\omega_0/2\pi$, 
and $\bar{n}=0.01$. Consequently, the initial thermal population 
beyond the ground state amounts to less than 1\%, corresponding to $k_BT\approx 0.217 \hbar\omega_0$ and we further have $\lambda=6.25\hbar\omega_0$.
The panels show $|\Delta E| = \hbar \omega_0$ (A), $|\Delta E| = 3.5\hbar \omega_0$ (B), $|\Delta E| = 5\hbar \omega_0$ (C), $|\Delta E| = 6.5\hbar \omega_0$ (D), 
and $|\Delta E| = 9\hbar \omega_0$ (E). (b) Dynamical evolution of the donor population (continuous lines), with exponential fits (dashed lines), for the parameters of panels A (yellow), B (pink), and C (violet). (c) 
Transfer rates (dark blue circles) extracted from the exponential fits in (b). 
The orange line displays the prediction (\ref{eq.Marcus-rate}) of the quantum nonadiabatic theory, with resonances of finite width determined by the cooling rate $\gamma$. 
}
\label{fig.quantum-nonadiabatic}
\end{figure}

\begin{figure*}
\centering
\includegraphics[width=\textwidth]{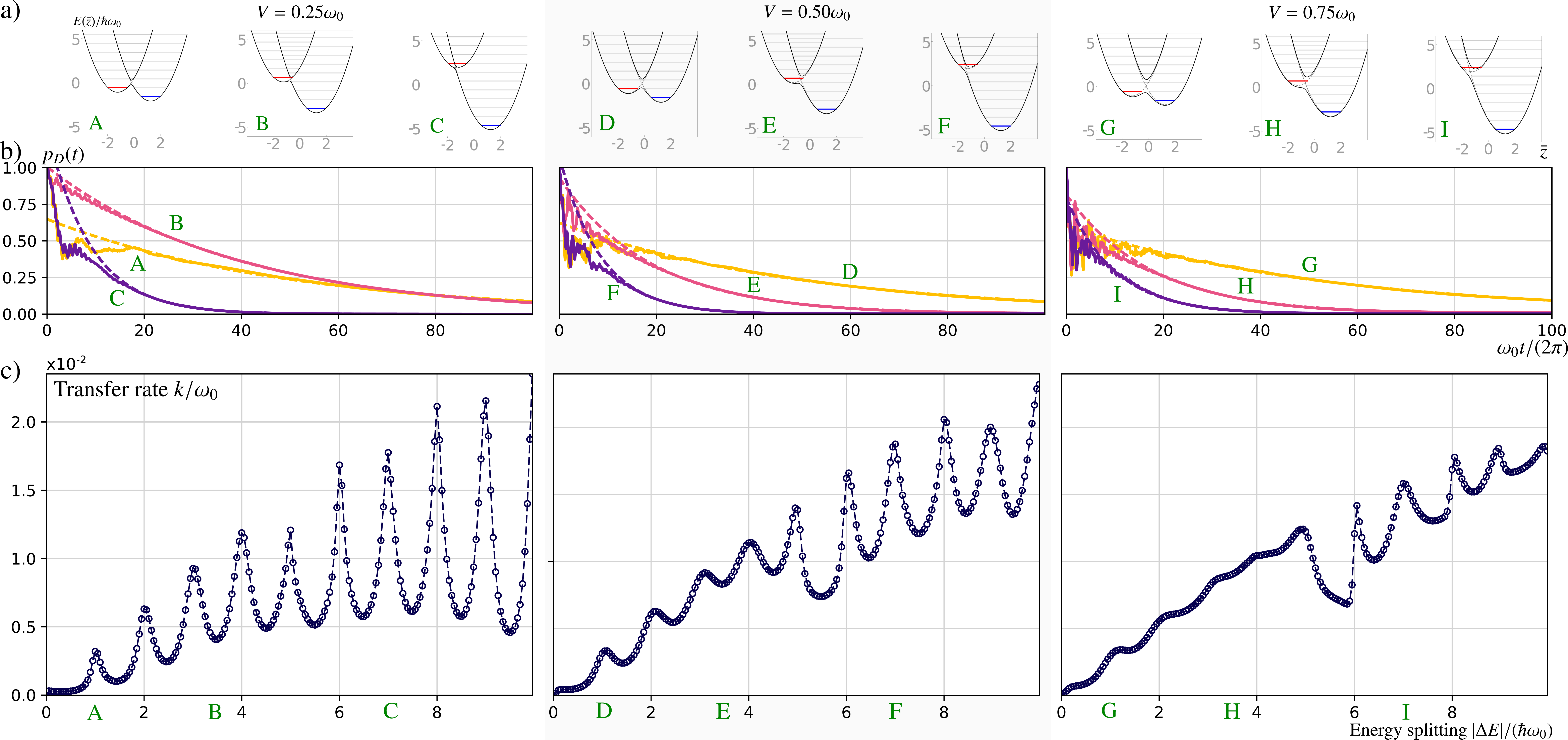}
\caption{ 
Crossover towards quantum adiabatic transfer and population trapping. 
(a) The Born-Oppenheimer surfaces split into lower and upper adiabatic surfaces, separated by an avoided crossing of size $2|V|$. The diabatic picture is no longer adequate due to the formation of hybridized vibronic states. (b) Time evolution of the donor state population as generated by Eqs.~(\ref{eq.H_tot}) and (\ref{eq.cooling-rate}), for $V=0.25\hbar\omega_0$ (left panels), $V=0.50\hbar\omega_0$ (center), and $V=0.75\hbar\omega_0$ (right). 
All remaining parameters are chosen as in Fig.~\ref{fig.quantum-nonadiabatic}. The plots show the evolution at $|\Delta E|=\hbar\omega_0$ (A,D,G; yellow lines), 
$|\Delta E|=3.5\hbar\omega_0$ (B,E,H; pink lines), and $|\Delta E|=7\hbar\omega_0$ (C,F,I; violet lines). The corresponding dashed lines show exponential fits after a transient 
evolution time of $\omega_0 t/2\pi = 20$. (c) Transfer rates extracted from the exponential fits of the simulated quantum dynamics, as a function of the donor acceptor energy gap $\Delta E$.}
 \label{fig.quantum-adiabatic}
\end{figure*}
Let us now turn to the discussion of quantum regimes for ET. When $k_B T \approx \hbar \omega_0$ the quantum nature of the vibrational mode has to be accounted for when modelling the transfer process. 
We first consider the parameter regime $|V| \ll \lambda/4$, where $V$ and $\hbar \gamma$ are of the same order of magnitude, which is described by quantum nonadiabatic (QN) ET theory \cite{marcusreview, kuzul,maykuhn,Nitzan}. 
In Fig.~\ref{fig.setupnew}~(a), this situation is sketched by indicating the respective energies $E_{D,\nu}$ and $E_{A,\nu'}$ of the uncoupled donor and acceptor vibronic eigenstates $\vert D \rangle \vert \nu \rangle$ and $\vert A \rangle \vert \nu' \rangle$ as horizontal lines. 
Population transfer is then facilitated by resonances between pairs of such states, as expressed by Fermi's golden rule \cite{marcusreview, kuzul,maykuhn,Nitzan}: 
\begin{align}
k_{\rm QN} &= \frac{2\pi}{\hbar} \vert V \vert^2 \sum_{\nu, \nu'} p_{\nu} \mathrm{FC}_{\nu, \nu'}  \; \delta (E_{D,\nu} - E_{A,\nu'}),
\label{eq.Marcus-rate}
\end{align}
where the $p_{\nu}$ describe the initial thermally distributed phonon populations and 
$\mathrm{FC}_{\nu, \nu'}  = \vert \langle \nu \vert e^{g (a^{\dagger} - a) /  \omega_0 } \vert \nu' \rangle \vert^2$ is the 
Franck-Condon factor between two displaced harmonic oscillators of identical frequency.

This ET transfer regime can be simulated on the ion-trap platform with standard parameters, i.e., by cooling to smaller phonon numbers as compared to those discussed in the classical regime above. 
Figure~\ref{fig.quantum-nonadiabatic}~(a) shows the vibronic energy levels in the quantum nonadiabatic regime for different values of $\Delta E$. The time evolution of the donor state population is fitted to an exponential decay [after a transient evolution of $\omega_0 t/2\pi=100$], see Fig.~\ref{fig.quantum-nonadiabatic}~(b), for $\bar{n}=0.01$ ($k_BT\simeq 0.217\hbar\omega_0$).
Figure~\ref{fig.quantum-nonadiabatic}~(c) depicts the variation of the obtained transfer rates with $\Delta E$. The classical Marcus parabola is now heavily modulated by vibronic resonances which allow for fast ET at sharply defined values of the donor-acceptor energy splitting. The transfer rates are extremely well described by the predictions of nonadiabatic theory, Eq.~(\ref{eq.Marcus-rate}), which is plotted for comparison \cite{footnoteEq7}. We note that this vibronic modulation of the resonances, which is readily accessible in an ion-trap quantum simulation, is washed out for molecular ET experiments in condensed-phase environments by uncontrolled disorder and additional coupling to low-frequency vibrations.
Figures~\ref{fig.cl-nonAd-regime} and~\ref{fig.quantum-nonadiabatic} demonstrate that current ion-trap technology can simulate well-known ET rate regimes. As such, these results define a benchmark for the experimental ability to simulate the ET Hamiltonian. 

With increasing electronic coupling $V$ (and all other parameters fixed) -- which is controlled by the injected laser power -- the above picture of resonant transitions between diabatic states is washed out by the opening of an appreciable avoided crossing between the donor and acceptor BO surfaces [see Fig.~\ref{fig.quantum-adiabatic} (a)], ultimately defining two new, excited and ground state adiabatic potential surfaces which no longer intersect. 

This trend towards a dominant role of $V$ manifests in a marked change of the dependence of the transfer rate on the energy gap: As one enters the regime $|V| \gg \hbar \gamma$, strong direct coupling $V$ quickly transfers initial population from the donor to the acceptor -- where it can only remain if the phonons created in the processes are promptly removed by dissipation
(implemented by the cooling laser with rate $\gamma$). 
In the above limit, however, population is transferred back and forth between donor and acceptor many times before dissipation renders the process irreversible. 
Consequently, $\gamma$ becomes the limiting factor which controls the transfer rate, and a simple model restricted to only the initial state $\vert D \rangle \vert 0 \rangle$ (corresponding to the the zero-temperature limit) and a single resonant vibronic state $\vert A \rangle \vert \nu \rangle$ yields a transfer rate of (see Appendix~\ref{app.QAd})
\begin{align}
k^{0\nu}_{\rm QA} \propto \nu \gamma, \label{eq:limit}
\end{align}
which is simply the decay rate of the $\nu$th vibrational level, according to the dissipator~(\ref{eq.cooling-rate}).
This regime of transfer rates limited by $\gamma$, where the direct coupling $V$ no longer affects the transfer rate as in Eq.~(\ref{eq.Marcus-rate}), is not readily accessible in a condensed phase environment where vibrational relaxation is fast.
Equation~(\ref{eq:limit}) neglects coupling to off-resonant states, which becomes important when the direct coupling $V$ approaches the phonon energy $\hbar\omega_0$ (leading to avoided crossings with widths on the order of phonon energies), as well as the effect of $V$ on the phonon lifetime due to hybridisation of vibrational levels near resonance. 
Nevertheless, it allows us to understand the basic transport mechanism as mediated by the vibrational excitations of the emerging adiabatic BO ground state -- which can therefore be considered as 
a quantum adiabatic (QA) transfer regime.

\begin{figure}
\centering
\includegraphics[width=0.42\textwidth]{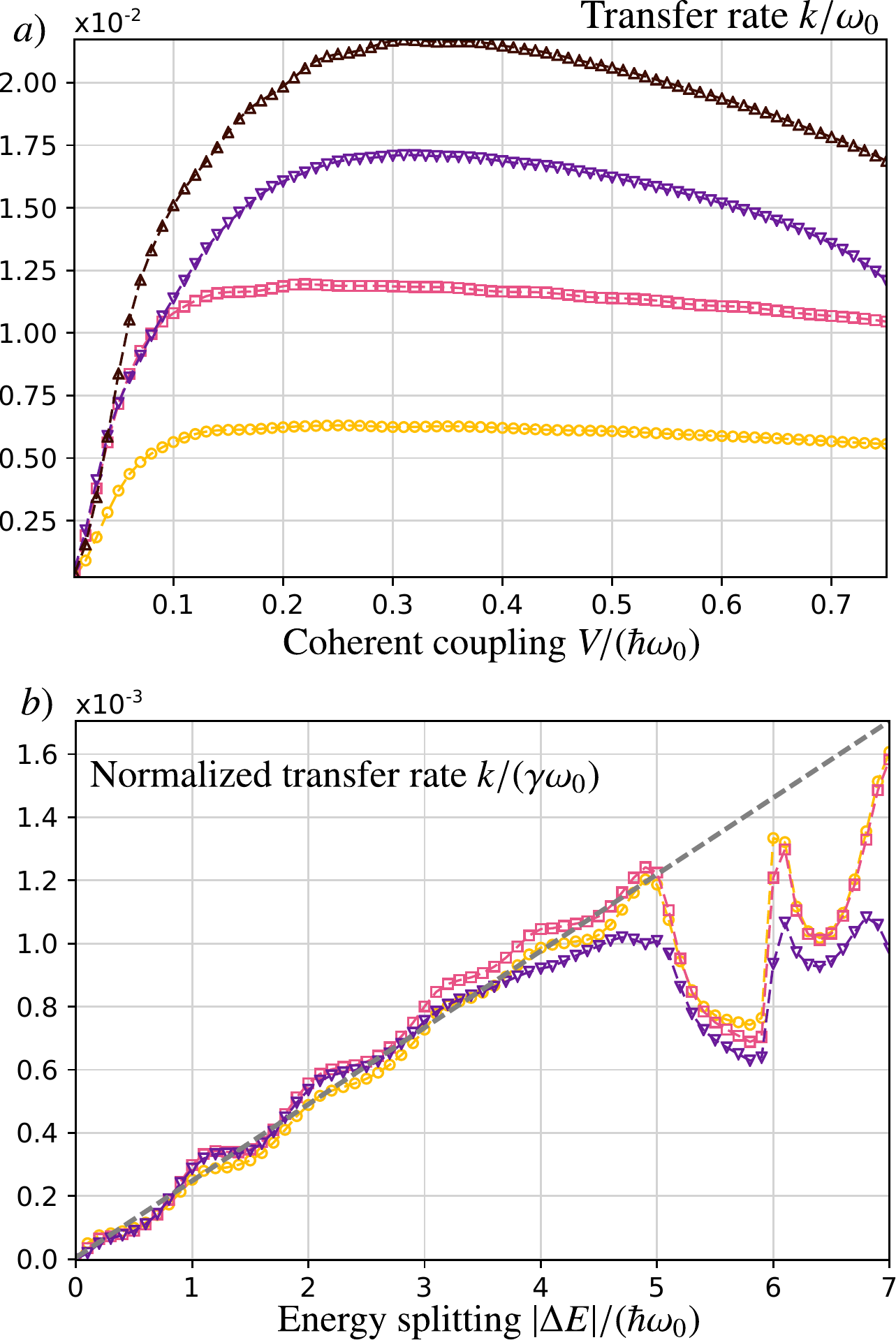}
\caption{  
Properties of the transfer rate in the quantum adiabatic limit. (a) Quantum ET rates as a function of the electronic coupling $V$ for $|\Delta E| = 2 \hbar \omega_0$ (yellow circles), 
$|\Delta E| =4 \hbar \omega_0$ (pink squares), $|\Delta E| =6 \hbar \omega_0$ (violet lower triangles) and $|\Delta E| =8 \hbar \omega_0$ (black upper triangles). 
 (b) Quantum ET rates divided by the cooling rate $\gamma$, as a function of the donor acceptor energy gap $\Delta E$, for $V=0.75\hbar\omega_0$ and $\gamma=0.025\omega_0/2\pi$
  (yellow circles), $\gamma=0.05\omega_0/2\pi$ (pink squares) and $\gamma=0.075\omega_0/2\pi$
 (violet lower triangles). The rates in the normal regime ($|\Delta E| <5 \hbar \omega_0$) approximately follow the quantum adiabatic limit~(\ref{eq:limit}), which predicts a linear dependence on $|\Delta E|$ (since the latter controls the resonantly coupled phonon number $\nu$). 
For $|\Delta E| > 5 \hbar \omega_0$, the process is no longer purely quantum adiabatic due to the trapping of part of the initial population in the upper adiabatic BO surface (see text). The dashed line shows a linear fit, $a|\Delta E|/(\hbar\omega_0)+b$, with $a=2.4\times 10^{-4}\gamma\omega_0$ and $b=5.2\times 10^{-6}\gamma\omega_0$. 
 All remaining parameters are chosen as in Fig.~\ref{fig.quantum-nonadiabatic}.
 }
 \label{fig.quantum-adiabatic2}
\end{figure}

We show examples of such quantum adiabatic time evolution in Fig.~\ref{fig.quantum-adiabatic}~(b), and the variation of the transfer rate as a function of the donor-acceptor energy gap $\Delta E$ in Fig.~\ref{fig.quantum-adiabatic}~(c), with the same parameters as in Fig.~\ref{fig.quantum-nonadiabatic}, except for the much larger electronic coupling 
strengths [$V/(\hbar\omega_0)=0.25,0.50,0.75$]. Whereas in Fig.~\ref{fig.quantum-nonadiabatic} the transfer rate is limited by the Franck-Condon factors of the transition 
from $\nu =0$ to $\nu'$ phonons, see Eq.~(\ref{eq.Marcus-rate}), and thus peaks at $|\Delta E| = 5 \hbar \omega_0$ (the classically activationless case), here the adiabatic rate is merely limited by the cooling rate, and on resonance keeps increasing with $\Delta E$. In the normal transfer regime ($|\Delta E| < 5 \hbar \omega_0$) the resonances broaden with increasing direct coupling $V$, and the transfer rates eventually converge to the adiabatic limit~(\ref{eq:limit}), see yellow and pink lines in Fig.~\ref{fig.quantum-adiabatic2}~(a). The driving-induced broadening of the 
resonances induces the rate to shrink at resonance as $V$ is further increased, while the off-resonant rates grow, which explains the slight decrease at very large values of $V$. 
Furthermore Fig.~\ref{fig.quantum-adiabatic2}~(b) demonstrates that $k/\gamma$ scales 
approximately linearly with $\nu$, as predicted by Eq.~(\ref{eq:limit}). 

Yet another structural feature comes into play in the inverted regime ($|\Delta E| > 5 \hbar \omega_0$): As apparent in particular in the limit of large 
$V$ [$\geq 0.50\omega_0$ in Fig.~\ref{fig.quantum-adiabatic}~(c)], the transfer resonances broaden less strongly, and, incidentally, sharpen asymmetrically. Furthermore, the transfer rates do not saturate with increasing $V$, as to be expected from Eq.~(\ref{eq:limit}), but instead decay after reaching a maximum 
for an optimal value of the coupling, see Fig.~\ref{fig.quantum-adiabatic2}~(a). 
Closer inspection of the spectral structure of the Hamiltonian~(\ref{eq.H_tot}) indeed reveals that the dominant eigenstate in the time evolution of the initial state selected here (approximately $\vert D \rangle \vert 0 \rangle$) is trapped and strongly localized in the excited adiabatic BO surface. In contrast to the population on the lower BO surface, the excited population remains trapped unless transfer is facilitated by a resonance, as we can clearly observe in Fig.~\ref{fig.quantum-adiabatic}~(c) for $V=0.75\omega_0$ (see Appendix~\ref{sec.width}). These observations imply that experimental control over the initial state, which can be achieved with high accuracy in trapped-ion experiments, defines yet another handle to control the ET efficiency in this largely unexplored parameter regime.

\section{Possible extensions of the setup}
\label{sec:extensions}
In this manuscript, we have studied ET simulations involving only a single vibrational mode. In many molecular ET reactions in condensed-phase environments a large number $N_{\rm vib}$ of vibrational modes $\mu$ are ET-active with $\lambda_{\mu}/\hbar \omega_{\mu} > 1$ and $k_B T / \hbar \omega_{\mu} \geq 1$. Quantum effects play a relevant role in the dynamics of such modes, but their exact treatment is usually intractable~\cite{footnoteintract}. Instead, numerical ET simulations use approximations which treat a large fraction of ET-active modes as classical vibrations. Trapped-ion quantum simulations provide an alternative to large-scale numerical simulations and semiclassical approximations, by reproducing the fully quantum ET dynamics in a well-controlled setup. Experimental quantum simulations of ET may thus complement theoretical, computational and experimental efforts on molecular ET systems, and may be used to benchmark approximate models. This opens up new pathways towards the microscopic understanding of ET processes, the identification of novel transfer regimes, and the design of highly efficient ET systems.

Possible extensions of our proposal include coherent couplings to multiple motional modes to enable vibronic transfer on a more complex vibrational backbone, or to study the influence of intermediate bridge states \cite{advchem2}. This could be implemented by increasing the number of ions and inducing laser couplings that encompass multiple motional modes~\cite{Govinda}. More flexibility in tuning the electronic transfer properties of the system is offered by engineered electronic couplings between different ions with tunable interaction range, which can be controlled via an effective spin-spin coupling using additional motional modes~\cite{MolmerSorensen,PorrasCirac}. These methods can be combined with two-dimensional traps with particular lattice geometries \cite{Hakelberg,KieferPRL2019}. Other studies may target the influence of different initial states, including nonequilibrium or even nonclassical vibrations \cite{molphys,Poyatos,Meekhof,Wolf,Home2019}. 

\section{Conclusions}
\label{sec:conclusions}
We have shown that trapped-ion analog quantum simulators provide a versatile testbed for studying phonon-mediated electron transfer under well-controlled, adjustable conditions. Numerical simulations demonstrate that a rather simple system composed of two ions coupled to a single phononic mode is able to reproduce essential features of donor-acceptor molecular ET. Continuously tuning the parameters allows us to connect and study a great variety of largely unexplored transport regimes which are not readily accessed by molecular ET experiments in a continuous way. When operated in a high-temperature regime, the predictions of Marcus theory can be verified for classical nonadiabatic ET with a high level of control over individual parameters. In the low-temperature regime, it is possible to study the emergence of quantum nonadiabatic vibronic transfer resonances. By further increasing the electronic coupling, quantum adiabatic transport is observed only in the normal regime, whereas the inverted regime gives rise to a transfer mechanism mediated by resonances due to trapped excited-state populations. It is impossible to realize such a controlled crossover between adiabatic and nonadiabatic regimes in molecular systems. 

Our work points out that dissipation and incoherent features in the long-time dynamics of quantum simulators, can give rise to interesting transport phenomena that are of high relevance, e.g., in molecular chemistry. These features become visible in the decay rates, after the coherent phenomena have largely been damped out by the engineered dissipation of phonons. Our approach is therefore complementary to predominantly coherent quantum simulations of the short-time dynamics that provide access to strongly entangled many-body systems by trying to mitigate decoherence~\cite{Schaetz12,Blatt12}. At longer time scales, the natural decoherence of the trapped-ion system becomes increasingly relevant. However, the time evolutions considered here are on the order of few milliseconds. This is still shorter than realistic quantum error correction cycles~\cite{Bermudez2017}, which in turn are subject to much stronger restrictions in terms of fidelity and coherence than our simulations.

\begin{acknowledgments} 
A.B., T.S. and S.S.S. thank FRIAS (Freiburg Institute of Advanced Studies) for support of the present cooperation through the Research Focus "Designed Quantum Transport in Complex Materials".
All authors thank A. Bermudez, H.-P. Breuer, and H. Haeffner for useful discussions.
F.~S. acknowledges funding from the European Research Council under the European Union's Seventh Framework Programme (FP7/2007-2013) Grant Agreement No. 319286 Q-MAC.  
M.G. acknowledges support by the LabEx ENS-ICFP: ANR-10-LABX-0010/ANR-10-IDEX-0001-02 PSL*.
T. S. acknowledges support from Deutsche Forschungsgemeinschaft [SCHA 973]. S.S.S. thanks the People Programme (Marie Curie Actions) of the European Union's Seventh Framework Programme (FP7/2007-2013), under the Research Executive Agency Grant 609305, and FRIAS for hosting him as an External Senior Fellow. The numerical simulations were performed with the aid of the \texttt{qutip} package for python \cite{qutip}.
\end{acknowledgments}

F.~S. and M.~G. contributed equally to this work.

\appendix

\section{Determination of the decay rate}
\label{app.decay-rate}
The information about the decay rates is contained in the long-time dynamics of the ion-trap system. In our simulations, these were obtained by propagating for sufficiently long times. Alternatively, if it is hard to extract the decay constant by an exponential fit (e.g., due to shorter propagation times), the decay rate can be efficiently approximated using \cite{Spiros92}
\begin{align}
k^{-1} &= \frac{ \int dt \; t p_D (t) }{\int dt \; p_D (t)}, \label{eq.k_DA}
\end{align}
where the integral excludes a transient initial time interval that is dominated by coherent effects. In the case of a pure exponential decay $\exp (- kt)$, the decay rate exactly reduces to the decay constant $k$. 

\section{Quantum adiabatic dynamics}
\label{app.QAd}
To understand the transition from the Fermi golden rule result~(\ref{eq.Marcus-rate}) to stronger interaction strengths $V$ in the quantum transport regime, we consider more closely the resonance between the initial state $\vert D \rangle \vert 0 \rangle$ and an acceptor state $\vert A \rangle \vert \nu \rangle$. Such a resonance is achieved for $\nu = |\Delta E | / (\hbar \omega_0)$ [see Fig.~\ref{fig.setupnew}~(a)]. The decay of the excited state can be described by the effective non-Hermitian Hamiltonian
\begin{align}
H &= \left( 
\begin{array}{cc}
0 & V \sqrt{ \mathrm{FC}_{0, \nu} } \\
V^{\ast} \sqrt{ \mathrm{FC}_{0 , \nu} } & - i \tilde{\gamma}
\end{array} 
\right), \label{eq.TLA}
\end{align}
where $\tilde{\gamma}$ is the decay rate of the acceptor state. Here, we assumed that the electronic coupling $V$ is further modulated by the Franck-Condon factor $\mathrm{FC}_{0, \nu}$ and neglected the coupling to other states. The donor population decays as
\begin{align}\label{eq:pdt}
p_D(t)=e^{- \tilde{\gamma} t}\left[\cosh\left(\frac{dt}{2}\right)+\frac{\tilde{\gamma}}{d}\sinh\left(\frac{dt}{2}\right)\right]^2,
\end{align}
with $d=\sqrt{\tilde{\gamma}^2-4|V|^2\mathrm{FC}_{0,\nu}}$. This can be mapped to the Jaynes-Cummings model on resonance, with a cavity mode that is coupled to a reservoir, leading to a Lorentzian spectral density \cite{Breuer}. Inserting Eq.~(\ref{eq:pdt}) into~(\ref{eq.k_DA}) yields the decay rate
\begin{align}
k^{0\nu}_{\rm QA}=\tilde{\gamma}\frac{1+\eta^2}{1+\frac{1}{2}\eta^4},
\end{align}
where the parameter $\eta=\tilde{\gamma}/(|V| \sqrt{\mathrm{FC}_{0,\nu}})$ determines the ratio of the phonon decay rate and the coherent coupling strength.

\subsection{Weak coupling}
The weak coupling limit, $|V|\sqrt{\mathrm{FC}_{0,\nu}}\ll\tilde{\gamma}$, is characterized by $\eta \gg 1$. We obtain
\begin{align}
k^{0\nu}_{\rm QA} = \frac{2 |V|^2 \mathrm{FC}_{0,\nu}}{\tilde{\gamma}} + \mathcal{O}(\tilde{\gamma}\eta^{-4}).
\end{align}
With $\tilde{\gamma} = \gamma$, this reproduces the Marcus result pertaining to the one transfer channel on resonance, and the delta-function broadened to a Lorentzian with width $\gamma$. It coincides with the nonadiabatic result~(\ref{eq.Marcus-rate}) on resonance.

\subsection{Strong coupling}
In the opposite limit, $\eta\ll 1$, we obtain
\begin{align}
k^{0\nu}_{\rm QA} = \tilde{\gamma}+\mathcal{O}(\tilde{\gamma}\eta^2),
\end{align}
leading to Eq.~(\ref{eq:limit}) for $\tilde{\gamma} = \nu \gamma$, i.e., the decay rate of the $\nu$th phononic level.

\subsection{Eigenstate width in the inverted regime}
\label{sec.width}
\begin{figure}[t]
\centering
\includegraphics[width=0.42\textwidth]{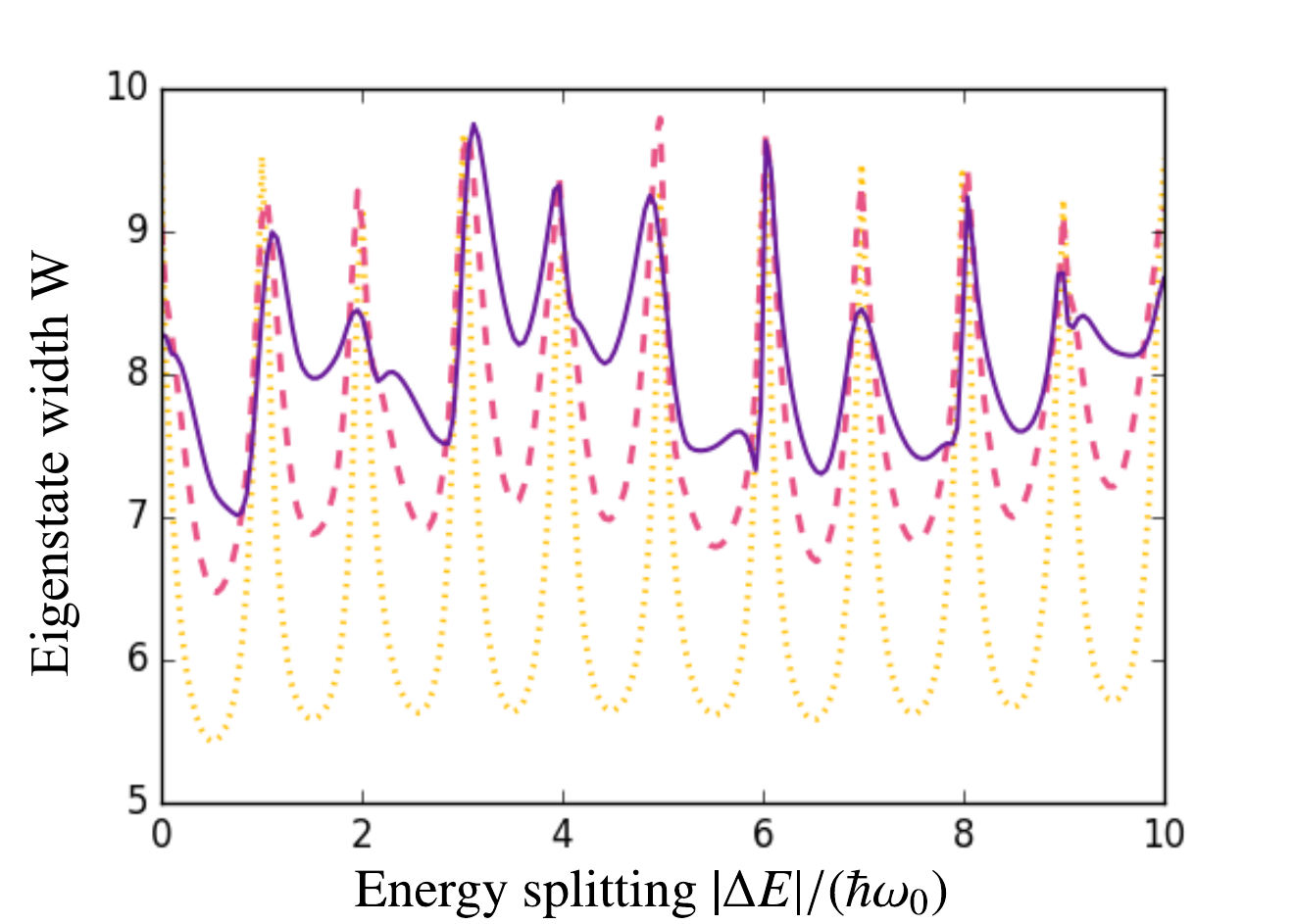}
\caption{  
Eigenstate width of the initial state according to Eq.~(\ref{eq.width}) for $V = 0.25 \hbar \omega_0$ (yellow, dot-dashed), $ 0.5 \hbar \omega_0$ (red, dashed), and $ 0.75 \hbar \omega_0$ (blue, solid line).
All remaining parameters are identical to Fig.~\ref{fig.quantum-adiabatic}.
 }
 \label{fig.eigenstate-width}
\end{figure}
To understand the survival of the sharp resonances in the inverted regime, we consider the ``eigenstate width" \cite{Bluemel87} of the initial state
\begin{align}
W &= \exp \left( - \sum_j p_j \ln p_j \right), \label{eq.width}
\end{align}
where $p_j \equiv \langle j \vert \rho \vert j \rangle$ is the overlap of the initial density matrix $\rho  = \vert D \rangle \langle D \vert \otimes \sum_{\nu} p_{\nu} \vert \nu \rangle \langle \nu \vert$ with the eigenstate $\vert j \rangle$  of the full Hamiltonian~(\ref{eq.H_tot}) and we use the convention $p \ln p = 0$ for $p = 0$. Intuitively, a larger value of $W$ indicates an increasingly complex dynamical evolution that, in a superposition of many different energy eigenstates, explores a larger fraction of the state space. This increases the likelihood of populating states that respond strongly to the laser cooling process and thus lead to an irreversible transfer of population from $\vert D \rangle \rightarrow \vert A \rangle$.

The width~(\ref{eq.width}) is shown in Fig.~\ref{fig.eigenstate-width} for three different interaction strengths $V$, corresponding to the three panels in Fig.~\ref{fig.quantum-adiabatic}. We observe a clear resonance structure that is less and less pronounced as the interaction strength is increased. 
However, the broadening arises asymmetrically: while the resonances are strongly washed out in the range $\Delta E / \hbar \omega_0 = 3, \dots, 5$, the eigenstate width remains more clearly peaked at integer values of $\Delta E / \hbar \omega_0$. In particular, the resonances at $\Delta E / \hbar \omega_0 = 6$ and $8$ remain very sharp, reflecting the behavior of the transfer rate in the simulation in Fig.~\ref{fig.quantum-adiabatic} (c).
Resonances at  $\Delta E / \hbar \omega_0 = 7$ and $9$ are broadened more strongly, again, in agreement with the transfer rate simulations. In summary, the analysis of the eigenstate width hence reveals (a) that in the inverted regime efficient transport $\vert D \rangle \rightarrow \vert A \rangle$ relies on the resonance of diabatic states, and (b) that the shape of the resonances is related to the distribution of energy eigenstates in the initial density matrix.


\begin{thebibliography}{99}
\bibitem{footnoteEq7} To account for the line broadening due to the sympathetic cooling, Eq.~(\ref{eq.cooling-rate}), we replace the delta function in Eq.~(\ref{eq.Marcus-rate}) by a normalized Lorentzian frequency distribution of width $2\pi\hbar\gamma$, i.e., $\delta (E_{D,\nu} - E_{A,\nu'}) \longrightarrow   2\hbar\gamma/((E_{D,\nu} - E_{A, \nu'})^2 + ( 2\pi\hbar\gamma)^2)$. The slight shift between the theory prediction and the simulation stems from an additional $\Delta E$-dependent broadening, which cannot be resolved in Fig.~\ref{fig.quantum-nonadiabatic} and which enhances the broadening of resonances in the inverted regime.
\bibitem{footnoteintract} In a fully quantum modeling, a very large number of phonon-excited states per mode ($n_{\mu} \gg \lambda_{\mu}/\hbar \omega_{\mu}$, $|\Delta E|/\hbar \omega_{\mu}$ and $k_B T / \hbar \omega_{\mu}$) would have to be included in the vibronic state space to achieve accuracy (i.e., $N_{\rm vib} \times n_{\mu}$ vibrational states for each electronic state).
\end{thebibliography}
\end{document}